\documentstyle[preprint,aps,pre,psfig]{revtex}

\begin{document}

\preprint{accepted to J. Phys. A}
\draft
\tighten

\title{ The three species monomer-monomer model 
in the reaction-controlled limit}

\author{Kevin E Bassler\footnote[1]{current address: Dept.\ of Physics,
University of Houston, Houston, TX} and Dana A Browne}

\address{
Department of Physics and Astronomy, Louisiana State University, Baton
Rouge, LA 70803, USA }

\maketitle{}

\begin{abstract}
We study the one dimensional three species monomer-monomer 
reaction model in the reaction controlled limit using mean-field theory
and dynamic Monte Carlo simulations.  The phase diagram consists of a
reactive steady state bordered by three equivalent adsorbing phases
where the surface is saturated with one monomer species.  The
transitions from the reactive phase are all continuous, while the
transitions between adsorbing phases are first-order.  Bicritical
points occur where the reactive phase simultaneously meets two
adsorbing phases.  The transitions from the reactive to an adsorbing
phase show directed percolation critical behaviour, while the universal
behaviour at the bicritical points is in the even branching annihilating
random walk class.  The results are contrasted and compared to previous
results for the adsorption-controlled limit of the same model.
\end{abstract}

\pacs{0570Ln, 8220Mj, 8265Jv, 6450Ht}

\section{Introduction}

Simple models with continuous phase transitions to an adsorbing steady
state where fluctuations are absent are prototypical models for
non-equilibrium critical phenomena.  These far-from-equilibrium models
arise in variety of contexts ranging from gravity-driven flow through a
porous medium to the spread of epidemics and to heterogeneous catalytic
reactions.

The critical behaviour at the transition in most of these models belongs
to the universality class of directed percolation (DP)\cite{DP}, which
is the simplest model with an adsorbing transition described by an
order parameter with no internal symmetry.  However, recently a number
of models with continuous adsorbing transitions having critical
properties not in the DP class have been
discovered\cite{BAW2,BAWe,otherDPmodels,3spec,BBB}.  These models, which
include branching annihilating random walks with even numbers of
offspring (BAWe)\cite{BAW2}, have similar critical properties and thus
form a distinct universality class.  The distinguishing feature of this
class is a conservation modulo 2 of the number of
defects\cite{CT,multicritDP}.

In one dimensional models, this critical behaviour can be seen in models
with two equivalent adsorbing states, where the defects are domain
walls between domains of the two different states.  The parity
conservation law, which requires that the number of defects always
remain either even or odd, arises naturally because two domain walls
are either created or destroyed each time a domain appears or
disappears.  The dynamical behaviour of these defects is very rich,
showing a variety of scaling behaviour some of which has no analogue in
the simpler DP universality class.

In order to study the role of adsorbing state symmetry in adsorbing
phase transitions, we recently introduced the three-species
monomer-monomer catalytic reaction model\cite{3spec}.  In this model,
three different species of monomers compete for lattice sites through
two fundamental dynamical processes:  (a) monomer adsorption onto empty
lattice sites, and (b) the annihilation reaction of two dissimilar
monomers adsorbed on nearest-neighbor lattice sites.  We previously
studied the one-dimensional version of the model in the
adsorption-controlled limit, where the reaction process happens
instantaneously.  We discovered a phase diagram consisting of three
adsorbing phases where the entire lattice is saturated by a single
monomer species, and a reactive phase. The phase transitions between
the reactive phase and the saturated phases are continuous and their
dynamical critical properties are those of the directed percolation
(DP) universality class. Phase transitions between saturated phases are
first-order. Of particular interest are {\em bicritical}
points\cite{BCP} where two saturated phases meet the reactive phase. At
those points the adsorbing state is two-fold degenerate and the
dynamical critical properties are those of the BAWe universality
class.  We also defined and measured three exponents associated with
the dynamic behaviour of an interface between domains of the two
equivalent adsorbing phases at the bicritical point, and we conjectured
that those exponents are universal, characteristic of the BAWe class.

In this paper we investigate the reaction-controlled limit of the
model, where the adsorption process happens instantaneously.  Using a
mean-field analysis and dynamical Monte Carlo simulations, we will show
that all of the qualitative features of the phase diagram remain the
same, and that all of the results for critical exponents are consistent
with those we found in the adsorption-controlled limit.  Most
importantly, we confirm our conjecture concerning the universal nature
of the interface dynamics exponents we defined previously.

In Section II we will present results from mean-field theory for this
model.  Section III discusses our results from dynamic Monte Carlo
calculations for the critical exponents of the model, studying the
critical behaviour of an isolated defect as well as the interface
between two different absorbing phases.  Our conclusions are presented
in Section IV.

\section{Mean Field Results}

The mean field theory we employ is based on a cluster expansion scheme
where the clusters consist of a chain of adjacent sites.  For each
allowed configuration of a cluster of a given size, we can write down
exact rate equations for the time evolution of the number of clusters
with that configuration.  For our problem, as we will see below, the
equations for a cluster of a given size require knowledge of the number
of clusters with one additional size, leading to an infinite chain of
equations.  The approximation results from truncating the chain of
equations at clusters of a given maximum size and replacing expressions
for the numbers of larger clusters with approximate forms.  This
technique, employed by Dickman\cite{DickmanMFT} for studying the ZGB
model\cite{ZGB1} for CO oxidation, is essentially the same as the
Kirkwood superposition approximation used in the theory of liquids.
This technique and its generalizations\cite{bAK} have been used
extensively in studies of lattice models.

For our model it is useful to describe a given configuration as having
``active'' bonds where the two adjacent sites are occupied by monomers
of different species, and ``inactive'' bonds where both sites are
occupied by the same species.  The configurations can change only where
there is an active bond.  Thus the evolution equation for a particular
cluster are found by properly enumerating all the ways an active bond
could produce the current cluster and also finding the number of ways
the current configuration could disappear because an active bond that
the cluster was a part of had changed.  For the purposes of deriving
the equations, it is easiest to account separately for the
disappearance and the production of active bonds.

If our system has $N$ sites, we define $N_A$ as the number of sites
occupied by $A$ monomers.  Similarly, the number of $AB$ pairs we
denote $N_{AB}$.  For the present we treat the number $N_{BA}$ of $BA$
pairs as distinct, although we expect that in the steady state the two
are equal.  Larger clusters are defined in an obvious fashion.

We begin with the equations for the clusters consisting of one site.
The fraction of $A$ sites can change if that site is either end of an
active bond.  The exact rate equation for the number of $A$ monomers is
\begin{eqnarray}
{d N_A\over d t}
 &= &2p_A(N_{BC}+N_{CB}+N_{AB}+N_{BA}+N_{AC}+N_{CA})\nonumber\\
    &&-(N_{AB}+N_{BA}+N_{AC}+N_{CA})
\end{eqnarray}
The first term on the right hand site represents the production of $A$
monomers at active bonds of any kind, the factor of two arising from
adding monomers at either end of the bond.  The second term represents
the rate that $A$ monomers disappear because they are part of an
active bond that is being updated.  The process of picking a site and
replacing an $A$ monomer with an $A$ monomer is divided between the two
terms.  The equations for $B$ and $C$ monomers are found by cyclicly
permuting the indices $A$, $B$ and $C$.

For the 2-clusters of pairs of sites, we proceed in a similar fashion.
The number of $AB$ clusters decreases every time the bond containing
$AB$ pair is chosen for update, or when the bond to either side of the
cluster is active and chosen for update.  Production of an $AB$ cluster
can proceed by updating an active bond to produce an $AB$ pair.  The
pair can also be produced by picking the bond to the right of the pair,
having the $A$ already present and supplying the $B$ in the update.
Similarly one can produce an $AB$ pair by having the $B$ present and
updating the bond to the left of the $A$ site to produce the $A$.  The
equation for $AB$ pairs is thus

\begin{eqnarray}
{d N_{AB}\over d t} &=& p_Ap_B (N_{AB}+N_{BA}+N_{AC}+N_{CA}+N_{BC}+N_{CB})
\nonumber\\
 && \mbox{} -N_{AB}\nonumber\\
 && \mbox{}+ p_B(N_{AAB}+N_{ABA}+N_{AAC}+N_{ACA}+N_{ABC}+N_{ACB})
\nonumber\\
  && \mbox{}- (N_{ABA} + N_{ABC}) \nonumber\\
 && \mbox{}+ p_A(N_{ABB}+N_{BAB}+N_{ACB}+N_{CAB}+N_{BCB}+N_{CBB})
\nonumber\\
  && \mbox{}- (N_{BAB}+N_{CAB})
\end{eqnarray}
If we were considering a pair occupied by the same monomer, like an
$AA$ pair, the process of selecting the bond occupied by the $AA$ pair
and updating it cannot occur, although the other steps in the $AB$ pair
equation can occur.  The equations for $AA$ pairs is
\begin{eqnarray}
{d N_{AA}\over d t} &=&
p_Ap_A (N_{AB}+N_{BA}+N_{AC}+N_{CA}+N_{BC}+N_{CB})\nonumber\\
 && \mbox{}+ p_A(N_{AAB}+N_{ABA}+N_{AAC}+N_{ACA}+N_{ABC}+N_{ACB})
\nonumber\\
  && \mbox{}- (N_{AAB} + N_{AAC}) \nonumber\\
 && \mbox{}+ p_A(N_{ABA}+N_{BAA}+N_{ACA}+N_{CAA}+N_{BCA}+N_{CBA})
\nonumber\\
  && \mbox{}- (N_{BAA}+N_{CAA})
\end{eqnarray}
The equations for larger clusters (of size $M$) are similar in
structure to these.  We change our notation slightly and denote
$N^{(M)}(\alpha_1\alpha_2\dots\alpha_M)$ as the number of clusters of
size $M$ consisting of the configuration
$\{\alpha_1\alpha_2\dots\alpha_M\}$ where each $\alpha_i$ is $A$, $B$,
or $C$.  The equation of motion for a size $M$ cluster is
\begin{eqnarray}
{d N^{(M)}(\alpha_1\dots\alpha_M)\over d t} &=&
\sum_{i=1}^{M-1} p_{\alpha_i}p_{\alpha_{i+1}}
\sum_{\beta\neq\gamma}
N^{(M)}(\alpha_1\dots\alpha_{i-1}\beta\gamma\alpha_{i+1}\dots\alpha_M)
\nonumber\\
 && \mbox{}- \sum_{i=1}^{M-1} (1-\delta_{\alpha_i,\alpha_{i+1}})
N^{(M)}(\alpha_1\dots\alpha_M)\nonumber\\
 && \mbox{}+p_{\alpha_1}\sum_{\beta\neq\gamma}
N^{(M+1)}(\beta\gamma\alpha_2\dots\alpha_M)
-\sum_{\beta\neq\alpha_1}
N^{(M+1)}(\beta\alpha_1\dots\alpha_M)\nonumber\\
 && \mbox{}+p_{\alpha_M}\sum_{\beta\neq\gamma}
N^{(M+1)}(\alpha_1\dots\alpha_{M-1}\beta\gamma)
-\sum_{\beta\neq\alpha_M}
N^{(M+1)}(\alpha_1\dots\alpha_M\beta)
\end{eqnarray}

\begin{figure}
\[ \psfig{file=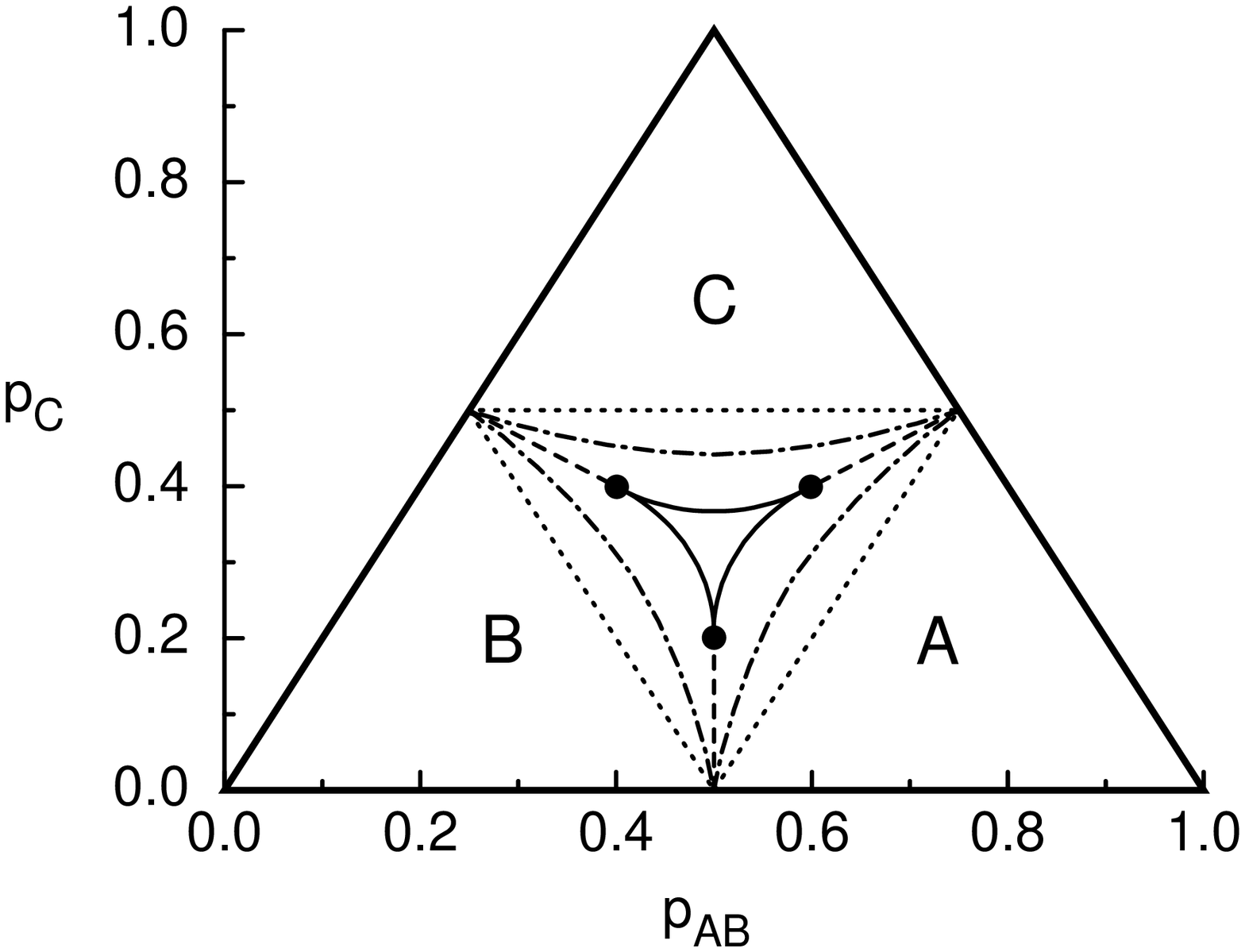,width=6.0in} \]
\caption{
Phase diagram showing three saturated phases (indicated by the
letters), and a reactive phase (the unlabeled center region). Solid
lines indicate continuous transitions.  Dashed lines indicate
first-order transitions. Bicritical points, shown as filled circles,
occur where two critical
lines meet a first-order line.  The site approximation is shown as a
dotted line and the pair approximation as a dash-dot line.}
\label{fig1}
\end{figure}

Despite the simple form of these equations and their close resemblance
to equations for two-component models that can be solved
exactly\cite{exact} and nearly exactly\cite{nearlyexact}, we have not
succeeded in solving the set of equations in closed form.  As
figure~\ref{fig1} shows, the Kirkwood superposition approximation for
clusters of size two and larger, where $N_{AB}=N_AN_B$, compares poorly
with the Monte Carlo results that will be presented below.  Improving
the approximation to keep the pairs of sites, and using
$N_{\alpha\beta\gamma}= N_{\alpha\beta}N_{\beta\gamma}/N_\beta$
improves the agreement somewhat.  However, even going to the level of
triples of sites, which was sufficient\cite{3spec} to get a qualitative
agreement with the Monte Carlo data in the adsorption--controlled
limit, failed to produce the coexistence line between the two saturated
phases.

The failure of this mean field approach to produce a
realistic position for the bicritical point is linked to the fact that
the probability of observing a long cluster of sites all filled with
one species, which is approximated in this cluster approach by the
probability of a smaller cluster raised to a power, will decay
exponentially with the size of the cluster.  However, the simulations
presented in this paper and previous work\cite{3spec} clearly
show that at the transitions large domains of each species are present
in the steady state, with the fundamental dynamical variables being the
domain walls.

Large domains by themselves do not necessarily cause the mean field
theory to fail, since a similar procedure applied to some monomer
annihilation models\cite{exact,nearlyexact} and cooperative sequential
adsorption models\cite{Coopadsorb} models yields exact results because
the cluster equations close.  However, the spatial correlations in
those models are very different from ours.   The adsorption models have
short-ranged correlations, while the monomer annihilation models have
long-range correlations that result from a simple linear diffusion
process.  In the present model, the long range correlations are
stronger than those produced by simple diffusion, so mean field theory
based on small clusters is not as successful here.  This lack of convergence
of mean field models has been seen in other 1D lattice models\cite{bAK}.
There it was ascribed to having the same phase diagram in all dimensions,
which is clearly a restatement of the assumed range of correlations.

\section{Simulations}

To study the critical properties of the three species monomer-monomer
model we used time-dependent Monte Carlo simulations. This well
established method is a form of ``epidemic'' analysis\cite{DMCS} in
which the average time evolution of a particular configuration that is
very close to an adsorbing state (defect dynamics), or very close to a
minimal width interface between two different adsorbing states
(interface dynamics), is measured by simulating a large number of
independent realizations.  Using this technique we located critical and
bicritical points and determined the universality classes of the
transitions.

In the reaction-controlled limit we are studying, reactions occur only
between nearest neighbor pairs of dissimilar monomers, that is only
between monomers connected by an active bond.  Near the continuous
phase transitions the number of active bonds is very small. Thus, a
traditional Monte Carlo algorithm that picks a bond at random and
attempts to react the monomers is inefficient because most bonds chosen
are inactive and no change occurs.  Instead, we use a variable time
algorithm in which a bond is randomly picked from a list of active
bonds, thereby assuring that the reaction will occur.  The reacting
pair of monomers immediately desorb and then two new monomers
immediately adsorb in their places.  The species of the new monomers
are chosen randomly according to the relative adsorption rates
$\{p_{\alpha}\}$.  The time elapsed during a step is $1/n_b(t)$ where
$n_b(t)$ is the total number of active bonds at that time.  Thus, on
average there is one attempted adsorption per bond, or per lattice
site, per unit time. We always start with a lattice large enough that
the active region never reaches a boundary; it is effectively an
infinite lattice.

During the simulations at the critical point we measured the survival
probability $P(t)$, i.e.\ the probability that the system had not
reached an adsorbing state by time $t$, the average number of active
bonds per run $\langle{n}_b(t)\rangle$, and the average mean-square
size of the active region per surviving run $\langle{R^2}(t)\rangle$.
At a continuous adsorbing phase transition as $t\rightarrow\infty$
these dynamical quantities show power law behaviour
\begin{equation}
P(t) \sim t^{-\delta} , \qquad
\langle{n}_b(t)\rangle \sim t^{\eta} , \qquad
\langle{R^2}(t)\rangle \sim t^{z} .
\label{quantities}
\end{equation}
The critical exponents $\delta$, $\eta$, and $z$
characterize the universality class of the phase transition.

Precise estimates of the location of the critical point and of the
exponents can be made by examining the local slopes of the curves of
the measured quantities on a log-log plot.  For example, the effective
exponent $\delta(t)$ defined as
\begin{equation}
-\delta(t) = \{ \ln\left[ P(t)/P(t/b) \right] / \ln b \}\,,
\label{localslopes}
\end{equation}
is a numerical approximation of the local slope of the survival
probability.  In our numerical studies we take $b=5$.  Similar
expressions define $\eta(t)$ and $z(t)$.  At the critical point, a
graph of the local slope versus $t^{-1}$ should extrapolate as
$t^{-1}\to 0$ to the critical exponent.  The correction to scaling is
expected to be linear in $t^{-1}$\cite{ctsnote}.  Away from the
critical point, the local slope should curve away from the critical
point value as $t^{-1}\to 0$.

\begin{figure}
\[ \psfig{file=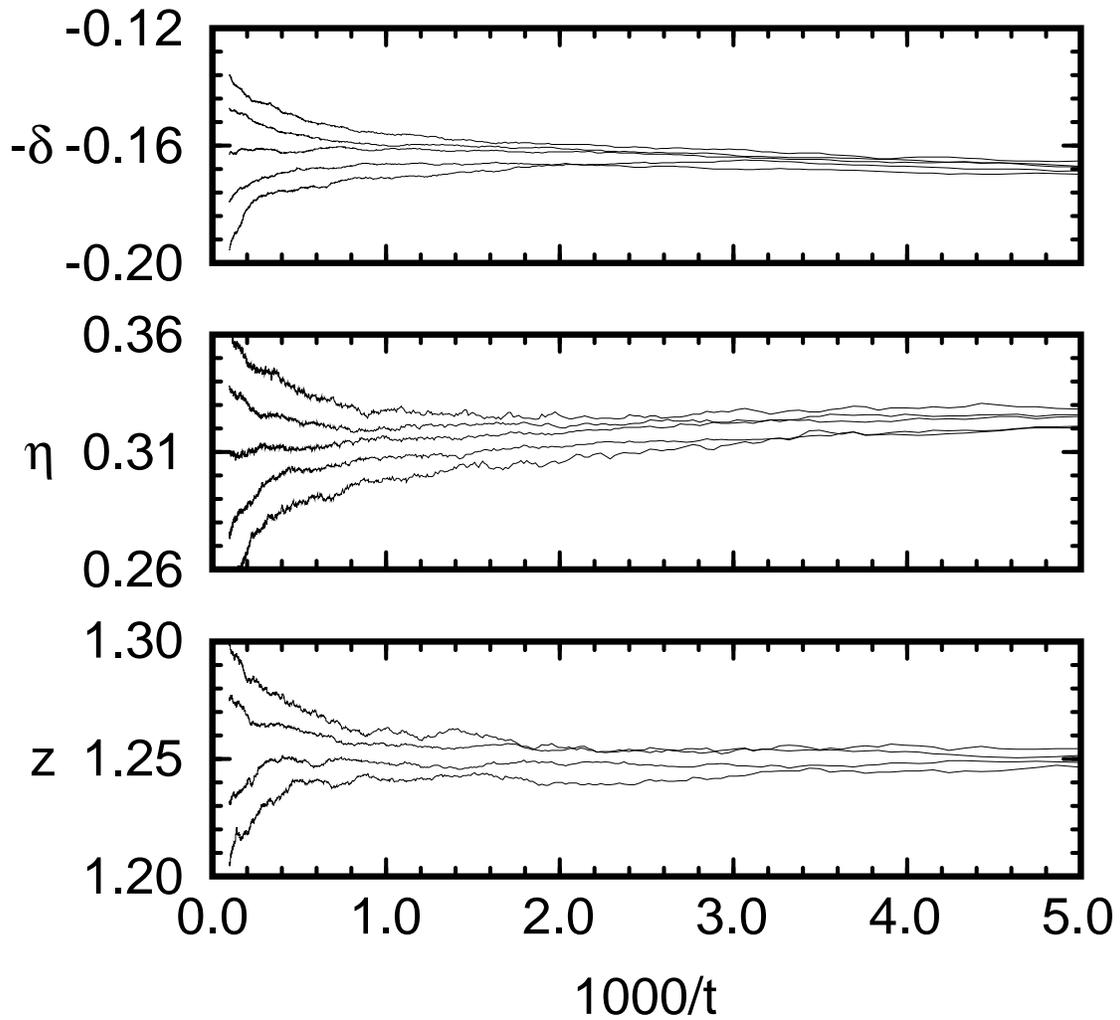,width=6.0in} \]
\caption{
Effective exponents using equation~(\protect\ref{localslopes}) with $b=5$
for the defect dynamics near the critical point at $p_{AB}=0.5$ on the
line where the C poisoned phase meets the reactive phase.  From top to
bottom, the 5 curves in each panel correspond to $p_C=0.3671$,
0.3672, 0.3673, 0.3674, and 0.3675, with the middle curve corresponding
to the critical point.}

\label{figcpls}
\end{figure}

The results of our simulations near the phase transition to the $C$
saturated phase at $p_{AB}=0.5$ are shown in figure~\ref{figcpls}.  These
data were extracted from $4 \times 10^5$ independent runs of up to
$10^4$ time steps at each parameter value.  We find a critical $C$
monomer adsorption rate of $\tilde{p}_C=0.3673(1)$, and that the
critical exponents are $\delta=0.155(5)$, $\eta=0.310(5)$, and
$z=1.260(10)$.  These values are consistent with our expectation that
this transition should be in the DP universality class, for which the
exponents are\cite{DPexp} $\delta=0.1596(4)$, $\eta=0.3137(10)$, and
$z=1.2660(14)$.  Our values for the exponents also satisfy the scaling
law\cite{DMCS} $4\delta + 2\eta=z$ to within the quoted errors.

The values for all of the exponents we find are slightly lower than
those found from series expansion studies\cite{DPexp}.  None of the
dynamic Monte Carlo studies in the literature examining the critical
exponents using local slope analysis include any discussion of how the
uncertainties in the exponents are determined.  Also, the errors quoted
are often
considerably smaller than ours even though less data were used.  We
surmise that the most common error determination involves an eyeball
estimate based on the scatter in the local slope curve, or by least
squares fits to the curve.  However, error estimates derived from least
squares fits to the local slope cannot be relied on because of the
strong correlation between values of the slope at later times on the
values at earlier times.  Also, there is an uncontrolled systematic
error if the critical point value is not known precisely.

We have therefore paid close attention to determining the error.
We divided our data into 10 equal sets of $4\times10^4$
trials.  With each data set we independently determined the exponent $z$
by two methods: a least squares fit to the local slope to extrapolate
the local slope to $t^{-1}=0$, and a linear least squares fit to plots
of $\ln R^2$ versus $t^{-1}$.  The results from each of the 10
independent data sets are combined to give unbiased estimates of the
mean and error for the exponent.

We found in this analysis that there was a substantial variation in the
value of $z$ found depending on whether data from short times are used.
Using the local slope data for times of $t>1000$, $t>5000$, and
$t>7500$ for the 10 data sets gave $z=1.256(1)$, $1.270(5)$,
$1.280(14)$ respectively.  While the short time data gave a very small
error, this is an artefact of using a large amount of data.  There
is clearly a systematic drift in the value (towards the series
expansion value) as the data from shorter times are dropped.  Similarly,
the least squares fits to $\ln R^2$ to a form $z\ln t + a/t + b$.  gave
$z=1.256(3)$ when fitting data for $t>1000$ and $z=1.272(5)$ when using
only data for $t>5000$.  Unfortunately the data for $t>5000$ span only
about 1/3 of a decade in time and so we cannot extrapolate further than
that.  The longer time data give values closer to that found by series
expansion methods, and in both methods of analysis we find
statistical errors that are considerably larger than those quoted in
other Monte Carlo studies that use less data and shorter times than we
adopted here.  This exercise points out the potential danger for
systematic error in this kind of dynamic Monte Carlo study.  The errors
we quote above for the exponents are derived from the $t>1000$ data sets
but have been expanded to account for the observed systematic drift we see.

We also studied the dynamics near the bicritical point at $p_{AB}=0.5$,
where the A-saturated and B-saturated phases coexist.  Since there are
two symmetric saturated phases, the dynamics at the bicritical points
are much richer than at critical points.  We used two distinct types of
epidemic analysis to study the dynamics there.

The first type of epidemic analysis we used to study the bicritical
dynamics is analogous to what was described above for the dynamics at
the critical point.  In this analysis, the average time evolution of a
configuration with a single defect in an otherwise adsorbing state was
measured.  Using an initial condition of a single B monomer in an
otherwise A-saturated phase, we measured the same three quantities as
before. In addition, we also measured the average number of B monomers
present $\langle n_B(t) \rangle$.  This quantity is expected to scale
in the limit $t\rightarrow \infty$ as $\langle n_B(t) \rangle \sim
t^{\eta_o}$ .  The characteristic exponent $\eta_o$ is not independent,
but is related to the other dynamic exponents by the scaling
law\cite{BBB}
\begin{equation}
\eta_o = z/2 - \delta .
\label{scalinglaw}
\end{equation}
Nevertheless, it is useful to measure $\eta_o$ as a numerical
check on the other exponents.

\begin{figure}
\[ \psfig{file=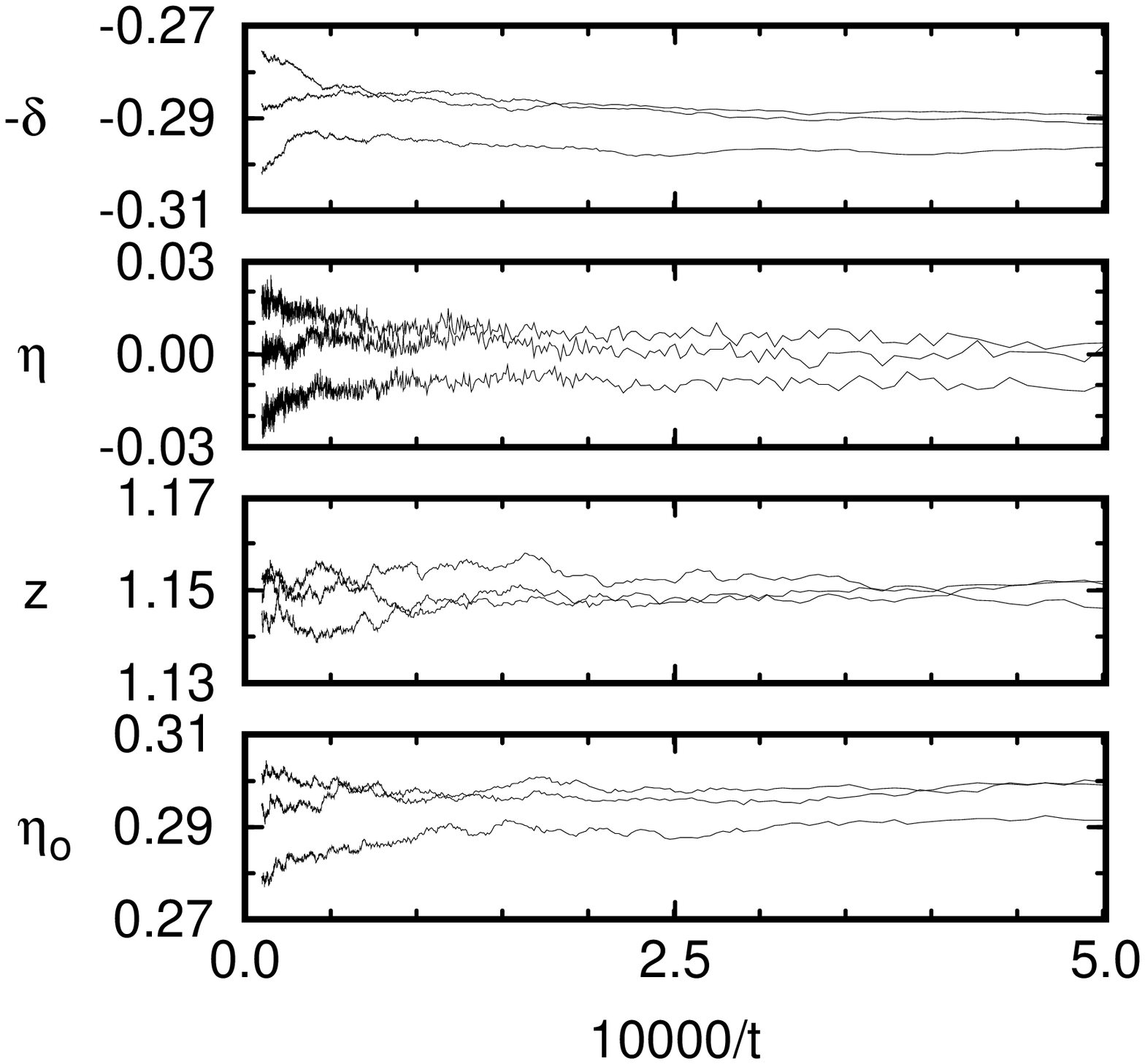,width=6.0in} \]
\caption{
Effective exponents for the defect dynamics near the bicritical point.
From bottom to top, the 3 curves in each panel correspond
to $p_C=0.201$, 0.2015, and 0.202, with the middle curve
corresponding to the bicritical point.}
\label{figbcpls}
\end{figure}

The local slope data for the four bicritical defect dynamics exponents
are shown in figure~\ref{figbcpls}.  These results were calculated using
$10^6$ runs of up to $10^5$ time steps each. The bicritical point is
located at $p_C=p_C^{*}=0.2015(5)$, and the exponents are $\delta=
0.285(5)$, $\eta=0.000(5)$, $z=1.15(1)$, and $\eta_o=0.29(1)$.  These
values satisfy the scaling law (\ref{scalinglaw}) and indicate that the
bicritical dynamics falls in the BAWe universality class, for which the
exponents are $\delta=0.285(2)$, $\eta=0.000(1)$, and
$z=1.141(2)$\cite{BAWe}.

The second type of epidemic analysis that we used to study the dynamics
at the bicritical point measures the average time evolution of a minimal
width interface between semi-infinite domains of the two different
adsorbing states\cite{3spec}.  The simulations are begun with a single
active bond joining domains of A and B monomers, and the trial is stopped
if the interface between the domains collapses back to a single active
bond.  We thus measure three dynamical quantities analogous to those
for the defect dynamics:  the probability that at time $t$ the
interface has not yet collapsed back to its minimum width, the number
of active bonds, and the mean square size of the interface.  These are
characterized at the bicritical point by the exponents $\delta'$,
$\eta'$ and $z'$ in analogy to equation~\ref{quantities}.

\begin{figure}
\[ \psfig{file=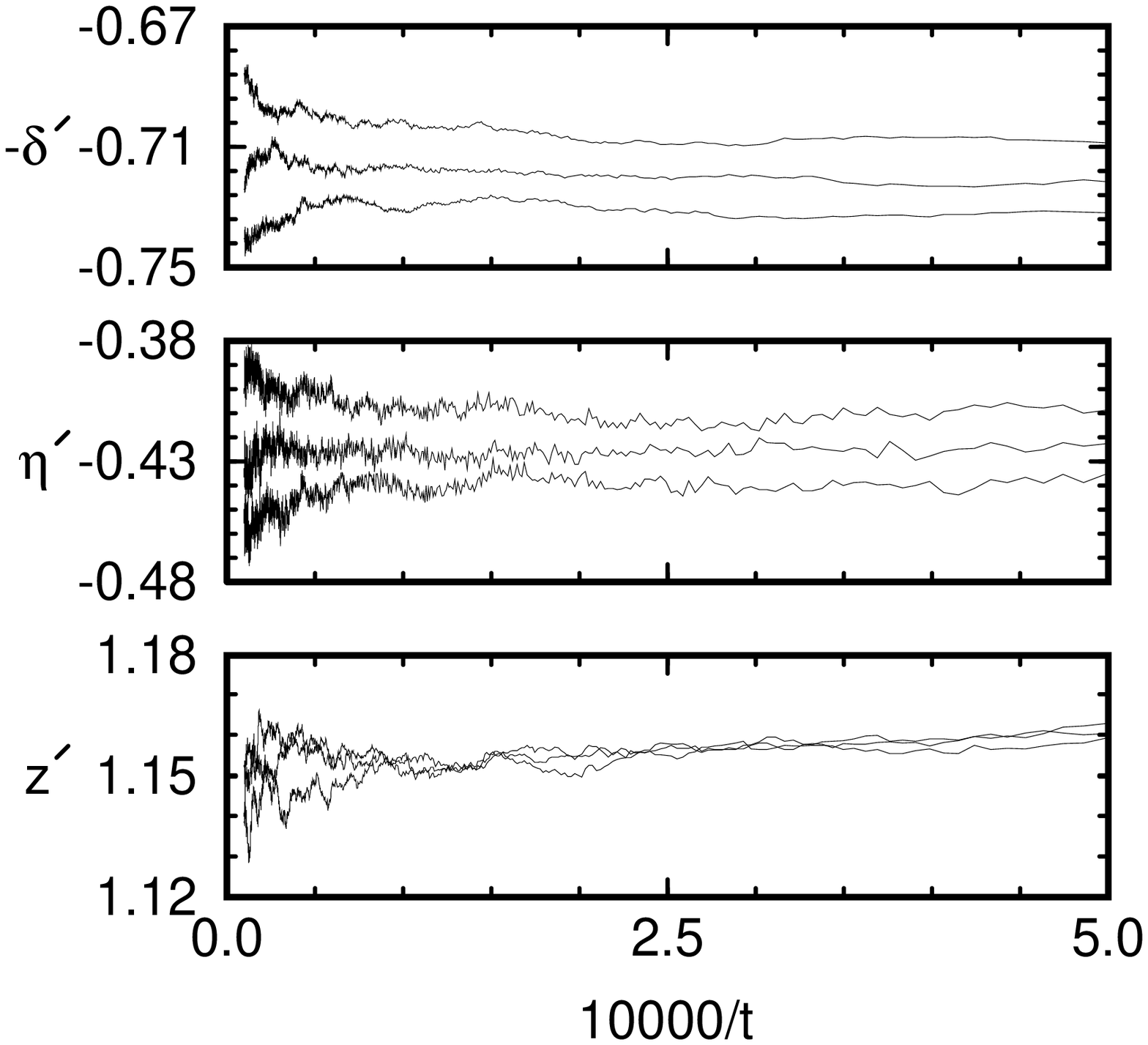,width=6.0in} \]
\caption{
Effective exponents for the interface dynamics near the bicritical
point with $b=5$.  From bottom to top, the 3 curves in each panel
correspond to $p_C=0.201$, 0.2015, and 0.202, with the middle curve
corresponding to the bicritical point.}
\label{figibcpls}
\end{figure}

Figure~\ref{figibcpls} shows the local slope results for these three
quantities. The slopes were derived from $10^7$ independent runs each
lasting up to $10^5$ time steps.  From these results we find
$\delta'=0.72(2)$, $\eta'=-0.43(2)$ and $z'=1.150(15)$. The values of
these exponents are consistent with those measured
previously\cite{3spec,BBB}, and confirm the conjecture\cite{3spec} that
these numbers are universal characteristics of the BAWe dynamics.
The value of the dynamic exponent $z$ or $z'$
describing the size of the active region during surviving runs, is
the same in both defect and interface dynamics simulations.
Furthermore, although the exponents $\delta$ and $\eta$ are different
in the two cases, their sums, which governs the time evolution of the
number of dissimilar pairs in just the surviving runs, are equal
\[
\delta + \eta=\delta'+\eta'\,.
\]
This shows that the critical spreading of the active region for models
with two symmetric adsorbing states is universal, independent of
whether defect or interface dynamics is being considered.  A similar
result holds for some one-dimensional systems with infinitely many
adsorbing states\cite{ManyAS}.

\section{Summary}

We have studied the one-dimensional three species monomer-monomer
reaction model in the reaction controlled limit.  As in our
study\cite{3spec} of this model in the adsorption-controlled limit, the
phase diagram consists of a reactive phase and three adsorbing phases
where the lattice is saturated with a single species.  The phase
transitions between the reactive phase and each of the saturated states
are continuous, but the transitions between different saturated phases
are first order.  Bicritical points exist where a first-order line
separating two adsorbing phases meets two critical lines separating the
reactive phase from each of those phases.

We constructed a mean-field cluster expansion of the model up through
triplets of sites.  The approximation correctly predicted the existence of
the four phases, but failed to predict that the bicritical points occur
in the interior of the phase diagram. This result differs from the
result of the same analysis for the adsorption-controlled limit of
the model. In that case\cite{3spec}, at the triplet approximation the
bicritical points moved off the edge of the phase diagram and the mean
field phase diagram became qualitatively correct. We presume that if
quadruples or possibly quintuples of sites are treated in the present
case, the bicritical points would move off the edge of the phase
diagram and the phase diagram would became qualitatively correct.

The dynamic critical behaviour at the transition between the reactive
phase and a poisoned phase is in the DP universality class, but at the
bicritical points, where there are two equivalent poisoned states, the
dynamic critical behaviour is in the BAWe class. Thus, the universality
class of the transition changes from DP to BAWe when the symmetry of
the adsorbing state is increased from one to two equivalent noiseless
states.  Furthermore, we have shown that having a symmetry in the
adsorbing states introduces a richness into the dynamics that is not
possible if there is a unique adsorbing state.  In particular, the
critical dynamics of the interfaces between two different adsorbing
states shows a sensitivity to how the dynamics is defined, and the
survival probability of fluctuations in the size of the interface from
its smallest value is described by a new universal exponent $\delta'$.
However, the critical spreading of the reactive region, be it a defect
in a single phase or a domain wall between phases, appears to be
insensitive to the choice of initial conditions.  This appears to
result from the fact that large reactive regions are insensitive to
whether the reactive regions are bounded by the same or different
saturated phases. We do not expect this result to be true in higher
dimensions where the entropy of domain walls can play a role and
non-universal critical spreading has been observed in other
models\cite{nonunivspreading}.

Thus, in conclusion, both the qualitative phase diagram and the
universality classes of the critical and bicritical points in the model
are equivalent in the reaction- and adsorption-controlled limits of the
three species monomer-monomer reaction model. We confirmed our previous
conjecture that the exponents describing the asymptotic behaviour of
those dynamics are universal numbers characteristic of the BAWe class.
Since these two limits correspond respectively to zero reaction rate
and infinite reaction rate, we predict that any finite reaction rate
version of the model will also be equivalent.

\section*{Acknowledgements}
This work was supported by the National Science Foundation
under Grant No.~DMR--9408634.  We wish to thank Giovanni Santostasi
for assistance in performing some of the simulations in this work.

\def\ZP{{\sl Zeit.\ Phys.}}
\def\PRL{{\sl Phys.\ Rev.\ Lett.\ }}
\def\PR{{\sl Phys.\ Rev.\ }}
\def\JPA{{\sl J. Phys.\ A: Math.\ and Gen.\ }}
\def\RMP{{\sl Rev.\ Mod.\ Phys.\ }}
\def\APNY{{\sl Ann.\ Phys.\ (New York)}}
\def\nonum{\par\item[]}

\begin{thebibliography}{99}
\bibitem{ctsnote}
Actually, this statement is known to be true only
for directed percolation \protect\cite{DP2} but it seems to also be
true\protect\cite{BAWe} for the BAWe universality class.
\end{thebibliography}
\end{document}